\documentclass{revtex4}
\usepackage{epsfig}
\usepackage{amssymb}
\usepackage{amsmath}
\usepackage{amsfonts}
\usepackage{graphicx}
\usepackage{mathrsfs}
\usepackage[dvips]{color}
\usepackage{multirow}

\newcommand{\fu}{\mathfrak{u}}

\newcommand{\fn}{{\mathfrak{n}}}

\newcommand{\fz}{\mathfrak{z}}

\newcommand{\bM}{\mathbf{M}}

\newcommand{\cB}{\mathcal{B}}

\newcommand{\cD}{\mathcal{D}}
\newcommand{\cH}{\mathcal{H}}
\newcommand{\cE}{\mathcal{E}}

\newcommand{\cO}{\mathcal{O}}
\newcommand{\cP}{\mathcal{P}}

\newcommand{\cT}{\mathcal{T}}
\newcommand{\cU}{\mathcal{U}}

\newcommand{\cX}{\mathcal{X}}

\newcommand{\cZ}{\mathcal{Z}}
\newcommand{\be}{\begin{equation}}
\newcommand{\ee}{\end{equation}}
\newcommand{\bea}{\begin{eqnarray}}
\newcommand{\eea}{\end{eqnarray}}
\newcommand{\nn}{\nonumber}
\newcommand{\kt}{\rangle}
\newcommand{\br}{\langle}

\newcommand{\ed}{\end{document}}

\newcommand{\bi}{\begin{itemize}}
\newcommand{\ei}{\end{itemize}}

\newcommand{\bce}{\begin{center}}
\newcommand{\ece}{\end{center}}

\newcommand{\sE}{\mathscr{E}}
\newcommand{\sF}{\mathscr{F}}

\newcommand{\sH}{\mathscr{H}}

\newcommand{\sT}{\mathscr{T}}
\newcommand{\RE}{{\rm Re}}
\newcommand{\IM}{{\rm Im}}

\begin{document}


\title{Lasing Threshold Condition for Oblique TE and TM Modes, Spectral Singularities,
and Coherent Perfect Absorption}

\author{Ali~Mostafazadeh and Mustafa~Sar{\i}saman}
\address{Departments of Mathematics and Physics, Ko\c{c}
University,\\ Sar{\i}yer 34450, Istanbul, Turkey\\
amostafazadeh@ku.edu.tr}

\begin{abstract}
We study spectral singularities and their application in determining the threshold gain coefficient  $g^{(E/M)}$ for oblique transverse electric/magnetic (TE/TM) modes of an infinite planar slab of homogenous optically active material. We show that $g^{(E)}$ is a monotonically decreasing function  of the incidence angle $\theta$ (measured with respect to the normal direction to the slab), while $g^{(M)}$ has a single maximum, $\theta_c$, where it takes an extremely large value.  We identify $\theta_c$ with the Brewster's angle and show that $g^{(E)}$ and $g^{(M)}$ coincide for $\theta=0$ (normal incidence), tend to zero as $\theta\to 90^\circ$, and satisfy $g^{(E)}<g^{(M)}$ for $0<\theta<90^\circ$. We therefore conclude that lasing and coherent perfect absorption are always more difficult to achieve for the oblique TM waves and that they are virtually impossible for the TM waves with $\theta\approx\theta_c$. We also give a detailed description of the behavior of the energy density and the Poynting vector for spectrally singular oblique TE and TM waves. This provides an explicit demonstration of the parity-invariance of these waves and shows that the energy density of a  spectrally singular TM wave with $\theta>\theta_c$ is smaller inside the gain region than outside it. The converse is true for the TM waves with $\theta<\theta_c$ and all TE waves.
\medskip

\noindent {Pacs numbers: 03.65.Nk, 42.25.Bs, 42.60.Da, 24.30.Gd}
\end{abstract}

\maketitle

\section{Introduction}

The history of modern physics includes happy instances where a purely mathematical concept turns out to describe an interesting physical phenomenon. A very recent example of such an instance is the remarkable observation that the spectral singularities of non-Hermitian differential operators \cite{naimark}, which constituted a subject of study in pure mathematics for over half a century \cite{ss-math}, turn out to find concrete physical realizations \cite{prl-2009,pra-2009,ss-phys-1} in terms of lasing at the threshold gain \cite{pra-2011a} and coherent perfect absorption (CPA) of electromagnetic waves \cite{CPA-longhi}.

Spectral singularities are certain points of the continuous spectrum of non-Hermitian operators. They draw the attention of mathematicians because they are responsible for a number of mathematical peculiarities that can never arise for Hermitian operators. In the context of the scattering theory, spectral singularities correspond to scattering states (with real and positive energy) that behave exactly like resonance states; they are zero-width resonances \cite{prl-2009}. This observation has provided ample motivation for the study of their physical aspects \cite{samsonov,ss-phys-2,pla-2011,jpa-2012,prsa-2012,ss-phys-3,pra-2013b,pra-2013d,ss-phys-4,ss-phys-5,aalipour} and nonlinear generalizations \cite{prl-2013,pra-2013c,liu,reddy,sap-2014}. For a recent review, see \cite{p123}.

An interesting outcome of the study of spectral singularities in optics is a mathematical derivation of the lasing threshold condition for a variety of optical setups. Among the notable applications of this approach is the determination of the threshold condition for the radial modes of single-layer and bilayer material with spherical geometry \cite{pla-2011,prsa-2012}, the TE modes of $\cP\cT$-symmetric and non-$\cP\cT$-symmetric bilayer slabs \cite{jpa-2012,sap-2014}, and the whispering gallery modes of cylindrical and spherical gain media \cite{pra-2013b,pra-2013d}.

\begin{figure}
    \begin{center}
    \includegraphics[scale=.40]{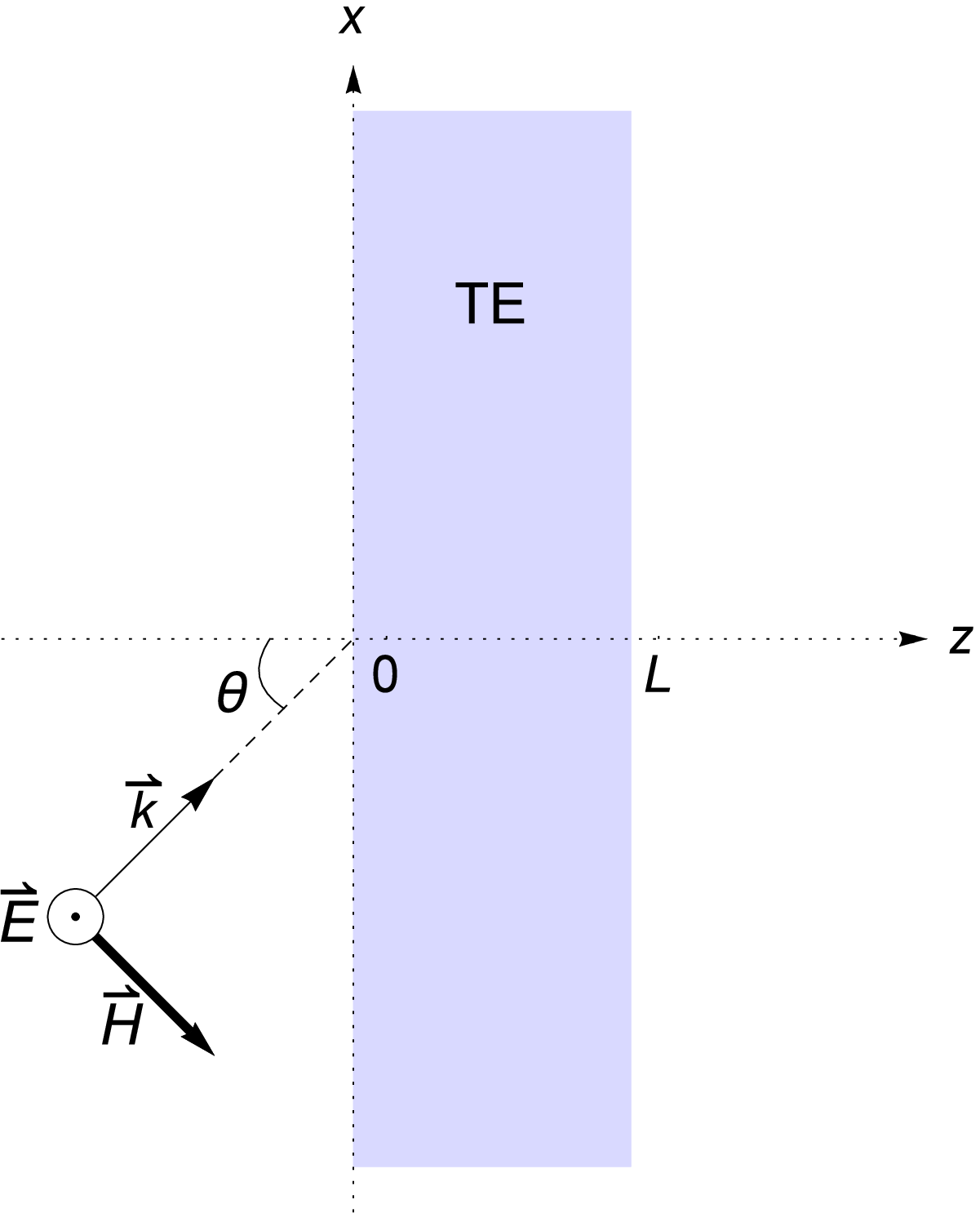}~~~
    \includegraphics[scale=.40]{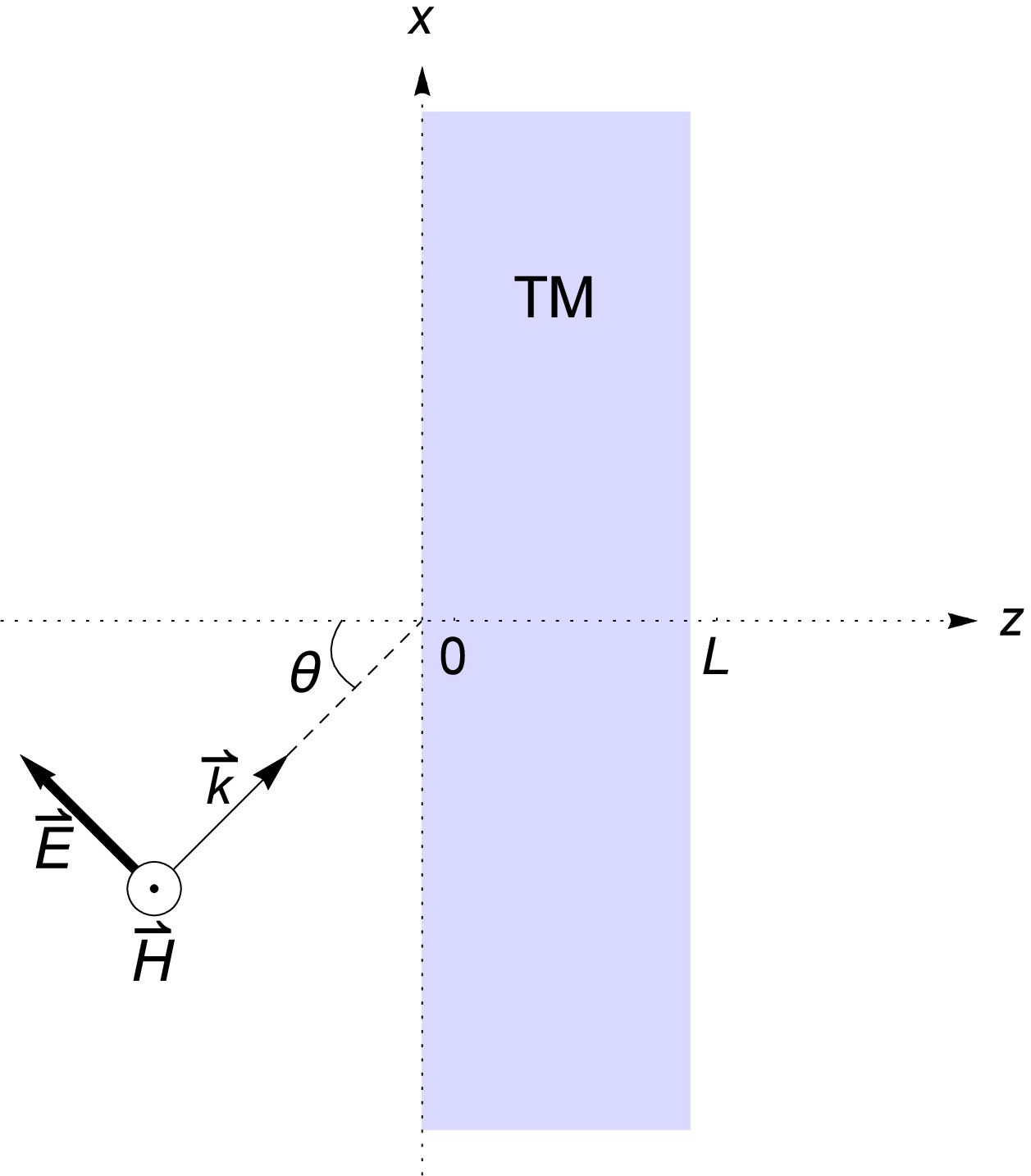}
    \caption{(Color online)  TE (on the left) and TM (on the right) modes of an infinite planar slab of thickness $L$.}
    \label{fig1}
    \end{center}
    \end{figure}

The fact that the mathematical condition for the emergence of a spectral singularity implies the laser threshold condition was initially shown in an explicit manner in \cite{pra-2011a} for the TE modes of an infinite planar slab laser where the wavevector $\vec k$ is taken along the normal direction $\hat e_z$ to the boundary of the slab. A natural extension of the approach of \cite{pra-2011a} is to consider the more general case where the direction of $\vec k$ deviations from $\hat e_z$, i.e., investigate the behavior of oblique waves (See Fig.~\ref{fig1}.)
Ref.~\cite{aalipour} treats this problem for the TE modes. The purpose of the present paper is to offer a more systematic and general solution of this problem that applies for both TE and TM modes. In particular, we show that much of the analysis of \cite{aalipour} can be reduced to that of \cite{pra-2011a} (i.e., to the case of normal incidence) by a simple change of variables, derive an explicit form of the laser threshold (and CPA) condition for oblique TE and TM modes, and examine the behavior of the Poynting vector and the energy density of the waves for the spectrally singular configurations of the system. This reveals a peculiar property of the spectrally singular TM waves having an emission angle larger than the Brewster's angle; the magnitude of the Poynting vector and the energy density of these waves are smaller in the gain region than in the surrounding vacuum.

\section{TE and TM Modes of an Infinite Planar Slab}

Consider an infinite planar slab of thickness $L$ that is aligned in the $x$-$y$ plane of some Cartesian coordinate system $\{(x,y,z)\}$. Let $\fn$ denote the complex refractive index of the content of the slab, and suppose that it is homogeneous and time-independent. Maxwell's equations describing the interaction of the electromagnetic waves with this system have the form:
    \begin{align}
    &\vec{\nabla}\cdot\vec{\cD} = 0, &&
    \vec{\nabla}\cdot\vec{\cB} = 0,
    \label{equation1}\\
    &\partial_t \vec{\cD}-\vec{\nabla} \times \vec{\cH}=\vec 0, &&
    \partial_t \vec{\cB}+\vec{\nabla} \times \vec{\cE}=\vec 0,
        \label{equation2}
    \end{align}
where $\vec\cE$ and $\vec\cH$ are the electric and magnetic fields,
    \begin{align*}
    \vec{\cD} := \varepsilon_0 \fz(z)\, \vec{\cE}, &&\vec{\cB}:=\mu_0\vec{\cH},
    \end{align*}
$\varepsilon_0$ and $\mu_0$ are respectively the permeability and permittivity of the vacuum, and
    \be
    \fz(z):=\left\{\begin{array}{ccc}
    \fn^2 & {\rm for} & 0\leq z\leq L,\\
    1 & {\rm for}  & z<0~{\rm and}~z>L.
    \end{array}\right.
    \label{e1}
    \ee
For time-harmonic electromagnetic fields, $\vec{\cE}(\vec r,t)=e^{-i\omega t}\vec{E}(\vec r)$ and $\vec{\cH}(\vec r,t)=e^{-i\omega t}\vec{H}(\vec r)$, Maxwell's equations reduce to
    \begin{align}
    &\left[\nabla^{2} +k^2\fz(z)\right] \vec{E}(\vec{r}) = 0, &&
    \vec{H}(\vec{r}) = -\frac{i}{k Z_{0}} \vec{\nabla} \times \vec{E}(\vec{r}),
    \label{equation4}\\
    &\left[\nabla^{2} +k^2\fz(z)\right] \vec{H}(\vec{r}) = 0,&&
    \vec{E}(\vec{r}) = \frac{i Z_{0}}{k \fz(z)} \vec{\nabla} \times \vec{H}(\vec{r}),
    \label{equation5}
    \end{align}
where $\vec r:=(x,y,z)$, $k:=\omega/c$ represents the wavenumber, and $c:=1/\sqrt{\mu_{0}\epsilon_{0}}$ and $Z_{0}:=\sqrt{\mu_{0}/\epsilon_{0}}$ are respectively the speed of light and impedance in vacuum.

The transverse electric (TE) and transverse magnetic (TM) waves correspond to the solutions of (\ref{equation4}) and (\ref{equation5}) for which $\vec E(\vec{r})$ and $\vec H(\vec{r})$ are respectively parallel to the surface of the slab. Let us choose our $y$-axis to be along this direction, and demand that outside the slab $\vec E(\vec{r})$ (respectively $\vec H(\vec{r})$) coincides with a plane wave with wavevector $\vec k$ in the $x$-$z$ plane, i.e.,
    \begin{align}
    &\vec k=k_x \hat e_x+ k_z \hat e_z, && k_x:=k\sin\theta, &&k_z:=k\cos\theta,
    \end{align}
where $\hat e_x,\hat e_y,$ and $\hat e_z,$ are respectively the unit vectors along the $x$-, $y$- and $z$-axes, and $\theta\in[-90^\circ,90^\circ]$ is the incidence angle (See Fig.~\ref{fig1}.) Then, the electric field for the TE waves and the magnetic field for the TM waves are respectively given by
    \begin{align}
    &\vec E(\vec{r})=\sE(z)e^{ik_{x}x}\hat e_y,
    && \vec H(\vec{r})=\sH (z)e^{ik_{x}x}\hat e_y,
    \label{ez1}
    \end{align}
where $\sE$ and $\sH$ are solutions of the Schr\"odinger equation
    \be
    -\psi''(z)+v(z)\psi(z)=k^2\psi(z),~~~~~~~~~~z\notin\{ 0,L\},
    \label{sch-eq}
    \ee
for the potential $v(z):=k^2[1+\sin^2\theta-\fz(z)]$. Because $v(z)$ is a piecewise constant potential, we can easily solve (\ref{sch-eq}) to obtain
    \be
    \psi(z):=\left\{\begin{array}{ccc}
    a_0\,e^{ik_z z} + b_0\,e^{-ik_z z} & {\rm for} & z<0,\\
    a_1\,e^{i{\tilde k}z} + b_1\,e^{-i{\tilde k}z} & {\rm for} & 0<z<L,\\
    a_2\, e^{ik_z z} + b_2\, e^{-ik_z z} & {\rm for} & z> L,
    \end{array}\right.
    \label{E-theta}
    \ee
where $a_i$ and $b_i$, with $i=0,1,2$, are complex coefficients, and
    \begin{align}
    &{\tilde k}:=k\sqrt{\fn^2-\sin^2\theta}=k_{z}\tilde\fn,
    &&\tilde\fn:=\frac{\sqrt{\fn^2 -\sin^2\theta}}{\cos\theta}.
    \label{tilde-parm}
    \end{align}
In particular, $\sE(z)$ and $\sH(z)$ are given by the right-hand side of (\ref{E-theta}) with generally different choices for the constants $a_i$ and $b_i$.

Substituting (\ref{ez1}) in the second equation in (\ref{equation4}) and (\ref{equation5}), we can find the magnetic field for the TE waves and the electric field for the TM waves. The resulting expressions are acceptable provided that they satisfy the appropriate boundary conditions for the problem, namely that at $z=0$ and $z=L$ the tangential components of $\vec E$ and $\vec H$ must be continuous functions of $z$. Table~\ref{table01} gives explicit expressions for the components of the electric and magnetic fields, and Table~\ref{table02} lists the corresponding boundary conditions.
They involve the following quantities.
    \bea
    \sF(z)&:=&\left\{\begin{array}{ccc}
    a_0\,e^{ik_{z}z} - b_0\,e^{-ik_{z}z} & {\rm for} & z<0,\\
    a_1\,e^{i{\tilde k}z} - b_1\,e^{-i{\tilde k}z} & {\rm for} & 0<z<L,\\
    a_2\, e^{ik_{z}z} - b_2\, e^{-ik_{z}z} & {\rm for} & z>L.
    \end{array}\right.
    \label{F-theta}\\
    \sT(x,z)&:=&\left\{\begin{array}{ccc}
    \sqrt{\fn^2-\sin^2\theta}\,e^{ik_{x}x}  & {\rm for} & z\in[0,L],\\
    \cos\theta\, e^{ik_{x}x}  & {\rm for} & z\notin[0,L],
    \end{array}\right.
    \label{T1-2}\\
    \fu&:=&\left\{\begin{array}{cc}
    \tilde \fn=\displaystyle\frac{\sqrt{\fn^2 -\sin^2\theta}}{\cos\theta} & \mbox{for TE waves},\\[6pt]
    \displaystyle\frac{\tilde\fn}{\fn^2}=
    \frac{\sqrt{\fn^2 -\sin^2\theta}}{\fn^2\cos\theta} & \mbox{for TM waves}.
    \end{array}\right.
    \label{u=}
    \eea
\begin{table}[!htbp]
    \begin{center}
	{
    \begin{tabular}{|c|c|}
    \hline
    TE-Fields & TM-Fields \\
    \hline & \\[-6pt]
    $\begin{aligned}
    & E_{x}=E_{z}=H_{y}=0\\[3pt]
    & E_{y}=\sE(z)\,e^{ik_{x}x}\\[3pt]
    &H_{x} =-\frac{\sF(z)}{Z_0}\,\sT(x,z)\\[3pt]
    &H_{z} =\frac{\sin\theta\, e^{ik_{x}x}\sE(z)}{Z_0}\\[3pt]
    \end{aligned}$ &
    $\begin{aligned}
    & E_{y}=H_{x}=H_{z}=0\\[2pt]
    & E_{x} =\frac{Z_{0}\,\sF(z)}{\fz(z)}\,\sT(x,z)\\[3pt]
    & E_{z} =- \frac{Z_{0}\sin\theta\, e^{ik_{x}x}\sH(z)}{\fz(z)}\\
    & H_{y} =\sH(z)\,e^{ik_{x}x}\\[-8pt]
    &
    \end{aligned}$\\
    \hline
    \end{tabular}}
    \vspace{6pt}
    \caption{Components of the TE and TM fields in cartesian coordinates. Here $\sE(z)$ is given by the right-hand side of (\ref{E-theta}), and $\sF(z)$ and $\sT(x,z)$ are respectively defined by (\ref{F-theta}) and (\ref{T1-2}).}
    \label{table01}
    \end{center}
\end{table}%
\begin{table}[!htbp]
    \begin{center}
	{
    \begin{tabular}{|c|c|}
    \hline
    &\\[-10pt]
    $z=0$ &
    $\begin{aligned}
    & a_0 + b_0 = a_1+ b_1, && b_0 - a_0 = \fu (b_1- a_1)\\[3pt]
    \end{aligned}$\\
    \hline
    &\\[-10pt]
    $z=L$ & $\begin{aligned}
    & a_1 e^{i{\tilde k}L} + b_1 e^{-i{\tilde k}L}= a_2 e^{ik_{z}L} + b_2 e^{-ik_{z}L} \\[3pt]
    & \fu (a_1 e^{i{\tilde k}L} - b_1 e^{-i{\tilde k}L})=
    a_2 e^{ik_{z}L} - b_2 e^{-ik_{z}L}
    \end{aligned}$\\[-8pt]
    &\\
    \hline
    \end{tabular}}
    \vspace{6pt}
    \caption{Boundary conditions for TE and TM waves. Here $\fu$ is defined by (\ref{u=}).}
    \label{table02}
    \end{center}
\end{table}

\section{Transfer Matrix and Spectral Singularities}

The transfer matrix for the system we consider is the $2\times 2$ matrix $\bM$ satisfying
    \be
    \left[\begin{array}{c}
     a_2 \\ b_2 \\ \end{array}\right]
    = \bM \left[\begin{array}{c}
            a_0 \\  b_0 \\ \end{array}\right].
    \label{M=}
    \ee
In view of the fact that the (left/right) complex reflection and transmission amplitudes, $R^{l/r}$ and $T^{l/r}$, are defined by
    \begin{align*}
    &\mbox{Left-incident waves~ ($b_2=0$):}~~~~ R^l:=\frac{b_0}{a_0},~~~~T^l:=\frac{a_2}{a_0},\\
    &\mbox{Right-incident waves ($a_0=0$):}~~~ R^r:=\frac{a_2}{b_2},~~~~T^l:=\frac{b_0}{b_2},
    \end{align*}
Equation~(\ref{M=}) relates the entries $M_{ij}$ of $\bM$ to $R^{l/r}$ and $T^{l/r}$ according to \cite{pra-2009}
    \begin{align}
    &R^l=-\frac{M_{21}}{M_{22}}, && R^r=\frac{M_{12}}{M_{22}}, && T^l=\frac{\det\bM}{M_{22}}, &&
    T^r=\frac{1}{M_{22}}.
    \label{RRT}
    \end{align}

With the help of the boundary conditions given in Table~\ref{table02} we can easily show that
    \begin{align}
    &M_{11} = \left\{ \cos({\tilde k}L)+\frac{i}{2}(\fu  +\fu^{-1}) \sin({\tilde k}L) \right\} e^{-ik_{z}L},
    \label{M11}\\
    &M_{12} = \frac{i}{2}(\fu - \fu^{-1}) \sin({\tilde k}L) e^{-ik_{z}L},
    \label{M12}\\
    &M_{21} = -\frac{i}{2}(\fu - \fu^{-1}) \sin({\tilde k}L) e^{ik_{z}L},
    \label{M21}\\
    &M_{22} = \left\{ \cos({\tilde k}L)-\frac{i}{2}(\fu +\fu^{-1}) \sin({\tilde k}L) \right\} e^{ik_{z}L}.
    \label{M22}
    \end{align}
In particular, we have
    \be
    \det\bM=1.
    \label{SL2}
    \ee
For TE waves the transfer matrix $\bM$ coincides with the one defined by the Schr\"odinger equation (\ref{sch-eq}) and the requirement that the solution be continuously differentiable at $z=0$ and $z=L$. In this case (\ref{SL2}) follows from well-known Wronskian identities of second order linear differential equations \cite{jpa-2009,sanchez-review}. These do not however apply for the TM waves, because TM waves are given by solutions of the Schr\"odinger equation (\ref{sch-eq}) involving jump conditions at $z=0$ and $z=L$. We can identify the latter with the effect of certain point interactions at these points. It is known that the transfer matrix for a point interaction does not generally satisfy (\ref{SL2}), \cite{jpa-2011}. The point interactions giving rise to the TM jump conditions at $z=0$ and $z=L$ are separately examples of the anomalous point interactions which violate (\ref{SL2}). However, one can show (using the results of \cite{jpa-2011}) that the contribution of each of these point interactions to $\det\bM$ cancels the other's, so that (\ref{SL2}) holds.

The spectral singularities correspond to the real and positive values of the wavenumber $k$ for which $M_{22}=0$. This implies that the reflection and transmission amplitudes diverge, \cite{prl-2009}. For the normally incident TE waves, where
    \begin{align}
    &\theta=0, && k_z=k, && {\tilde k}=k\fn, &&\fu=\tilde\fn=\fn,
    \label{normal}
    \end{align}
Equations~(\ref{M11}) -- (\ref{M22}) reproduce the known expression for the transfer matrix of a complex barrier potential, and $M_{22}=0$ takes the form  \cite{pra-2011a,pra-2013a}:
    \be
    e^{-2ikL\!\fn} = \left(\frac{\fn-1}{\fn+1}\right)^{\!\!2}.
    \label{normal-ss}
    \ee
In view of (\ref{M22}), we can obtain the spectral singularities in the oblique TE and TM modes of our system by making the following substitution in (\ref{normal-ss}).
    \begin{align}
    & k\to\frac{\tilde k}{\fn}=k\sqrt{1-\frac{\sin^2\theta}{\fn^2}}, &&\fn\to\fu=\left\{\begin{array}{cc}
    \tilde\fn & \mbox{for TE modes},\\[3pt]
    \displaystyle\frac{\tilde\fn}{\fn^2} & \mbox{for TM modes}.
    \end{array}\right.
    \label{substitute}
    \end{align}
This gives
    \be
    e^{-2i\tilde kL} = \left(\frac{\fu-1}{\fu+1}\right)^{\!\!2}.
    \label{normal-ss-u}
    \ee

Next, we substitute (\ref{tilde-parm}) and (\ref{u=}) in (\ref{normal-ss-u}) to obtain the following more explicit relation for the spectral singularities in oblique TE and TM modes of our model.
    \begin{align}
    &e^{-2i k L\sqrt{\fn^2-\sin^2\theta}}=
    \displaystyle\left(\frac{\sqrt{\fn^2-\sin^2\theta}-\fn^\ell\cos\theta}{
    \sqrt{\fn^2-\sin^2\theta}+\fn^\ell\cos\theta}\right)^{\!\!2},
    \label{TEM-ss}\\
    &\ell:=\left\{\begin{array}{cc}
    0 &  \mbox{for TE modes},\\
    2 &  \mbox{for TM modes}.
    \end{array}\right. \nn
    \end{align}
The expression given in (\ref{TEM-ss}) for the TE modes differs from the result reported in Ref.~\cite{aalipour}. The root of this discrepancy is that the author of \cite{aalipour} takes different wave vectors inside and outside the slab and relates their direction using the usual expression for Snell's law, namely $\fn\sin\theta=\fn'\sin\theta'$. Clearly, this cannot hold for complex refractive indices with different phases. We should also like to emphasize that a proper analysis of this problem should only rely on the use of Maxwell's equations and the boundary conditions on the electromagnetic fields. In particular, there is no need for invoking the use of consequences of the latter such as Snell's law and deal with the subtilises associated with their correct generalization and proper application. This is the basic strategy we pursue in the present article.

\section{Threshold Gain}

By definition the gain coefficient is given by
    \be
    g:=-2k\kappa=-\frac{4\pi\kappa}{\lambda},
    \label{gain-def}
    \ee
where $\kappa$ is imaginary part of the refractive index $\fn$ and $\lambda:=2\pi/k$ is the wavelength. The gain coefficient required for the emergence of a spectral singularity is known as the threshold gain coefficient \cite{pra-2011a}. For the oblique TE/TM modes we denote this quantity by $g^{(E/M)}$ and use (\ref{TEM-ss}) to derive an expression for it. This is known as the lasing threshold condition in optics, \cite{silfvast}.

Let us denote the real part of $\fn$ by $\eta$, so that
    \be
    \fn=\eta+i\kappa,
    \label{fn=}
    \ee
and take the logarithm of both side of (\ref{TEM-ss}) to obtain
    \be
    k=\frac{i}{2L\fn'} \ln\left(\frac{\fn'-\fn^\ell\cos\theta}{\fn'+\fn^\ell\cos\theta}\right)^{\!\!2},
    \label{TEM-ss-2}
    \ee
where $\fn':=\sqrt{\fn^2-\sin^2\theta}$. If we solve (\ref{gain-def}) for $k$ and substitute the result and (\ref{fn=}) in (\ref{TEM-ss-2}), we obtain a complex equation involving $g$, $\eta$, $\kappa$, and $\theta$. The left hand-side of this equation is real, so it should be equal to the real part of the right-hand side, and the imaginary part of the right-hand side should vanish. This gives a pair of real equations that can be manipulated to arrive at
    \bea
    g^{(E/M)}&=&\frac{2\,\IM(\fn)}{L\,\IM(\fn')}
    \ln\left|\frac{\fn'-\fn^\ell\cos\theta}{\fn'+\fn^\ell\cos\theta}\right|,
    \label{th-g=1}\\
    \lambda^{(E/M)}&=&\frac{2\pi L\,\RE(\fn')}{\pi m-\varphi^{(E/M)}},
    \label{th-lambda=1}
    \eea
where $\RE$ and $\IM$ respectively stand for the real and imaginary parts of their argument, $\lambda^{(E/M)}$ is the wavelength of the spectrally singular TE/TM wave generated for the gain coefficient $g^{(E/M)}$, $m$ is a positive integer (mode number), and $\varphi^{(E/M)}$ denotes the principle argument (phase angle) of $(\fn'-\fn^\ell\cos\theta)/(\fn'+\fn^\ell\cos\theta)$.

\begin{figure}
    \begin{center}
    \includegraphics[scale=.8]{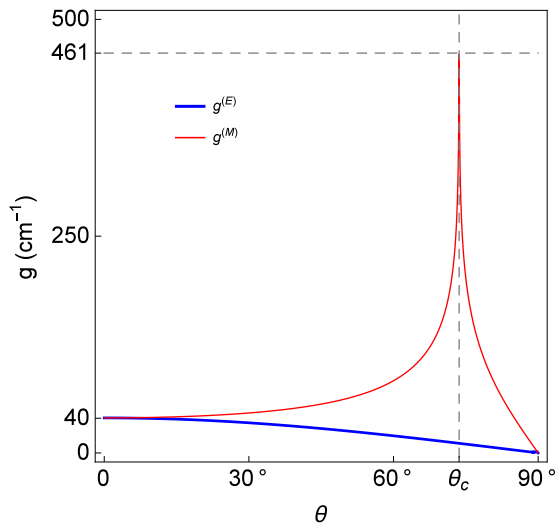}
    \caption{(Color online) Plots of the threshold gain coefficient $g$ as a function of $\theta$ for the TE waves (thick blue curve) and TM waves (thin red curve) of wavelength $1500\,{\rm nm}$ in a slab with $L=300~\mu{\rm m}$ and $\eta= 3.4$. The dotted vertical line marks the critical angle $\theta_c \approx 73.6105^{\circ}$ at which $g^{(M)}$ takes its maximum value, namely $461.113\,{\rm cm}^{-1}$.}
    \label{fig2}
    \end{center}
    \end{figure}

Equation~(\ref{th-g=1}) is the lasing threshold condition for oblique TE and TM modes. Setting $\ell=\theta=0$, it reproduces the result given in \cite{pra-2011a}. Figure~\ref{fig2} shows the plots of $g^{(E/M)}$ as a function of $\theta$ for a semiconductor gain medium with $L=300~\mu{\rm m}$ and $\eta=3.4$ that is obtained by setting $\lambda^{(E/M)}=1500\,{\rm nm}$ and using (\ref{gain-def}) and (\ref{th-g=1}).
The following are some of the properties $g^{(E/M)}$ that are demonstrated in Fig.~\ref{fig2}.
    \begin{enumerate}
    \item $g^{(E)}(0)=g^{(M)}(0)$, i.e., the slab begins lasing in the TE and TM waves propagating in the normal direction to the slab at the same value of the gain.
    \item For $0<\theta<90^\circ$, $g^{(E)}(\theta)<g^{(M)}(\theta)$. Therefore, it is more difficult to initiate lasing in the oblique TM modes than the oblique TE modes, i.e., it requires more pumping power.
    \item As $\theta\to 90^\circ$, both $g^{(E)}(\theta)$ and $g^{(E)}(\theta)$ tend to zero. This is to be expected, because for $\theta\to 90^\circ$ the optical path inside the gain region becomes infinitely long. The situation resembles that of spectrally singular whispering gallery modes \cite{pra-2013b,pra-2013d}. It corresponds to the infinite radius limit of the latter.
    \item $g^{(E)}$ is a smooth monotonically decreasing function of $\theta$, whereas  $g^{(M)}(\theta)$ increases for $\theta<\theta_c$ and deceases for $\theta>\theta_c$, for some critical angle $\theta_c=73.6015^\circ$.
    \end{enumerate}

In order to understand the nature of $\theta_c$ and the behavior depicted in Fig.~\ref{fig2}, we consider typical optically active material which satisfies $|\kappa|\ll \eta-1$. This allows for a perturbative treatment of (\ref{th-g=1}) and (\ref{th-lambda=1}). Expanding the right-hand side of these equations in a power series in $\kappa$ and ignoring quadratic and higher order terms, we find
    \bea
    g^{(E)}(\theta)&\approx&\frac{4\sqrt{\eta^2-\sin^2\theta}}{L\eta}\ln
    \frac{|\sqrt{\eta^2-\sin^2\theta}+\cos\theta|}{\sqrt{\eta^2-1}},
    \label{g-E=}\\
    \lambda^{(E)}&\approx&
    \frac{2L\eta}{m}\left[\sqrt{1-\frac{\sin^2\theta}{\eta^2}}+
    \frac{2\kappa\cos\theta}{\pi m (\eta^2-1)}\right],
    \label{L-E=}
    \eea
and
    \bea
    g^{(M)}(\theta)&\approx&\frac{2\sqrt{\eta^2-\sin^2\theta}}{L\eta}\ln\left|
    \frac{\sqrt{\eta^2-\sin^2\theta}+\eta^2\cos\theta}{
    \sqrt{\eta^2-\sin^2\theta}-\eta^2\cos\theta}\right|,
    \label{g-M=}\\
    \lambda^{(M)}&\approx&
    \frac{2L\eta}{m}\left[\sqrt{1-\frac{\sin^2\theta}{\eta^2}}+
    \frac{4\kappa\cos\theta[\eta^2-1+\cos(2\theta)]}{\pi m (\eta^2-1)[\eta^2-1+(\eta^2+1)\cos(2\theta)]}\right],
    \label{L-M=}
    \eea
where $\eta':=\sqrt{\eta^2-\sin^2\theta}$ and ``$\approx$'' stands for the fact that we have ignored quadratic and higher order terms in $\kappa$. Equations~(\ref{g-E=}) and (\ref{g-M=}) provide practically more appealing expressions for the lasing threshold (and CPA) condition (\ref{th-g=1}).

A closer examination of (\ref{g-M=}) shows that its right-hand side diverges whenever $\sin\theta=\eta/\sqrt{\eta^2+1}$, equivalently $\tan\theta=\eta$. This suggests that
    \be
    \theta_c\approx\tan^{-1}(\eta)=:\theta_b,
    \label{thetacritical}
    \ee
where $\theta_b$ denotes the Brewster's angle \cite{jackson} of the slab in the absence of gain. Indeed, the Brewster's angle $\theta_b$ for the sample used in Fig.~\ref{fig2} is $73.6015^\circ$. Therefore, our approximate expression (\ref{thetacritical}) for $\theta_c$ is in perfect agreement with the exact numerical result.

Equation (\ref{g-M=}) seems to imply that $g^{(M)}$ diverges for $\theta=\theta_b$. We should however recall that this equation is reliable only if $|\kappa|\ll\eta-1$. The prediction that $g^{(M)}$ takes extremely large values in the vicinity of $\theta_b$ conflicts with this condition, because $g$ is proportional to $\kappa$. In fact, it is easy to see from (\ref{th-g=1}) that $g^{(M)}$ is a bounded function for $\kappa\neq 0$; it attains a large but finite maximum value at $\theta=\theta_c$. For the sample used in Figure~\ref{fig2}, we find $g^{(M)}(\theta_c)=461.113~{\rm cm}^{-1}$, which is about an order of magnitude larger than the attainable gain coefficients for typical high-gain material \cite{silfvast}. Therefore, lasing (and CPA) are virtually unattainable for oblique TM modes with $\theta$ in the close vicinity of the Brewster's angle. This is actually quite easy to understand because for this modes the boundaries of the slab have very small reflectance (originating from the imaginary part of the complex refractive index.) This in turn reduces the internal reflections and the optical path inside the slab, and results in a very large threshold gain.

\begin{figure}
    \begin{center}
    \includegraphics[scale=.60]{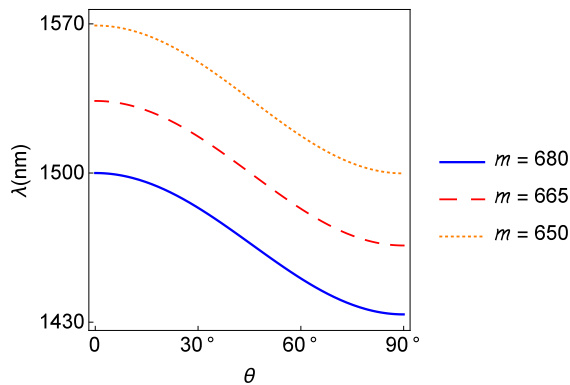}
    \caption{(Color online) Graphs of the wavelengths $\lambda^{(E/M)}$ as a function of $\theta$ for $L=300~\mu{\rm m}$, $\eta = 3.4$, $\kappa =-10^{-4}$, and $m=650, 665, 680$. The difference between $\lambda^{(E)}$ and $\lambda^{(M)}$ are too small to be visible.}
    \label{fig3}
    \end{center}
    \end{figure}

In typical situations, $L\gg\lambda$. Therefore, according to (\ref{L-E=}) and (\ref{L-M=}), the mode number $m$ is much greater than $1$. This in turn implies that, for the same mode number, the difference between $\lambda^{(E)}$ and $\lambda^{(M)}$ is negligibly small. Figure~\ref{fig3} shows the plots of $\lambda^{(E/M)}$ as a function of $\theta$ for the same gain medium as in Fig.~\ref{fig2} and different mode numbers $m$.

\section{Accounting for Dispersion}

The characterization of spectral singularities that we provide in the preceding section takes the refractive index $\fn$ and the wavenumber $k$ as independent parameters. In this section, we provide a description of the spectral singularities for a more realistic situation where we take into account the effect of dispersion.

Suppose that our slab is made out of a doped host medium of refraction index $n_0$ that we can model by a two-level atomic system with lower and upper level population densities $N_l$ and $N_u$, resonance frequency $\omega_0$, damping coefficient $\gamma$, and the dispersion relation
    \be
    \fn^2= n_0^2-
    \frac{\hat\omega_p^2}{\hat\omega^2-1+i\hat\gamma\,\hat\omega},
    \label{epsilon}
    \ee
where $\hat\omega:=\omega/\omega_0$, $\hat\gamma:=\gamma/\omega_0$, $\hat\omega_p:=(N_l-N_u)e^2/(m_e\varepsilon_0\omega_0^2)$, $e$ is electron's charge, and $m_e$ is its mass. We can express $\hat\omega_p^2$ in terms of the imaginary part $\kappa_0$ of $\fn$ at the resonance wavelength $\lambda_0:=2\pi c/\omega_0$ according to $\hat\omega_p^2=2n_0\hat\gamma\kappa_0+\cO(\kappa_0^2)$, \cite{pra-2011a}.
Inserting this relation in (\ref{epsilon}), using (\ref{fn=}), and neglecting quadratic and higher order terms in $\kappa_0$, we obtain \cite{pla-2011}
    \begin{align}
    &\eta\approx n_0+\kappa_0f_1(\hat\omega),
    &&\kappa\approx\kappa_0f_2(\hat\omega),
    \label{eqz301}
    \end{align}
where
    \begin{align}
    & f_1(\hat\omega):=\frac{\hat\gamma(1-\hat\omega^2)}{(1-\hat\omega^2)^2+
    \hat\gamma^2\hat\omega^2}, &&
    f_2(\hat\omega):=\frac{\hat\gamma^2\hat\omega}{(1-\hat\omega^2)^2+
    \hat\gamma^2\hat\omega^2}.\nn
    \end{align}
In view of (\ref{gain-def}), we also have $\kappa_0=-\lambda_{0}g_0/4\pi$. Substituting this equation in (\ref{eqz301}) and using the resulting relations together with (\ref{gain-def}) in (\ref{TEM-ss}) we can determine the $\lambda$ and $g_{0}$ values for the spectral singularities.

\begin{figure}
    \begin{center}
    \includegraphics[scale=.50]{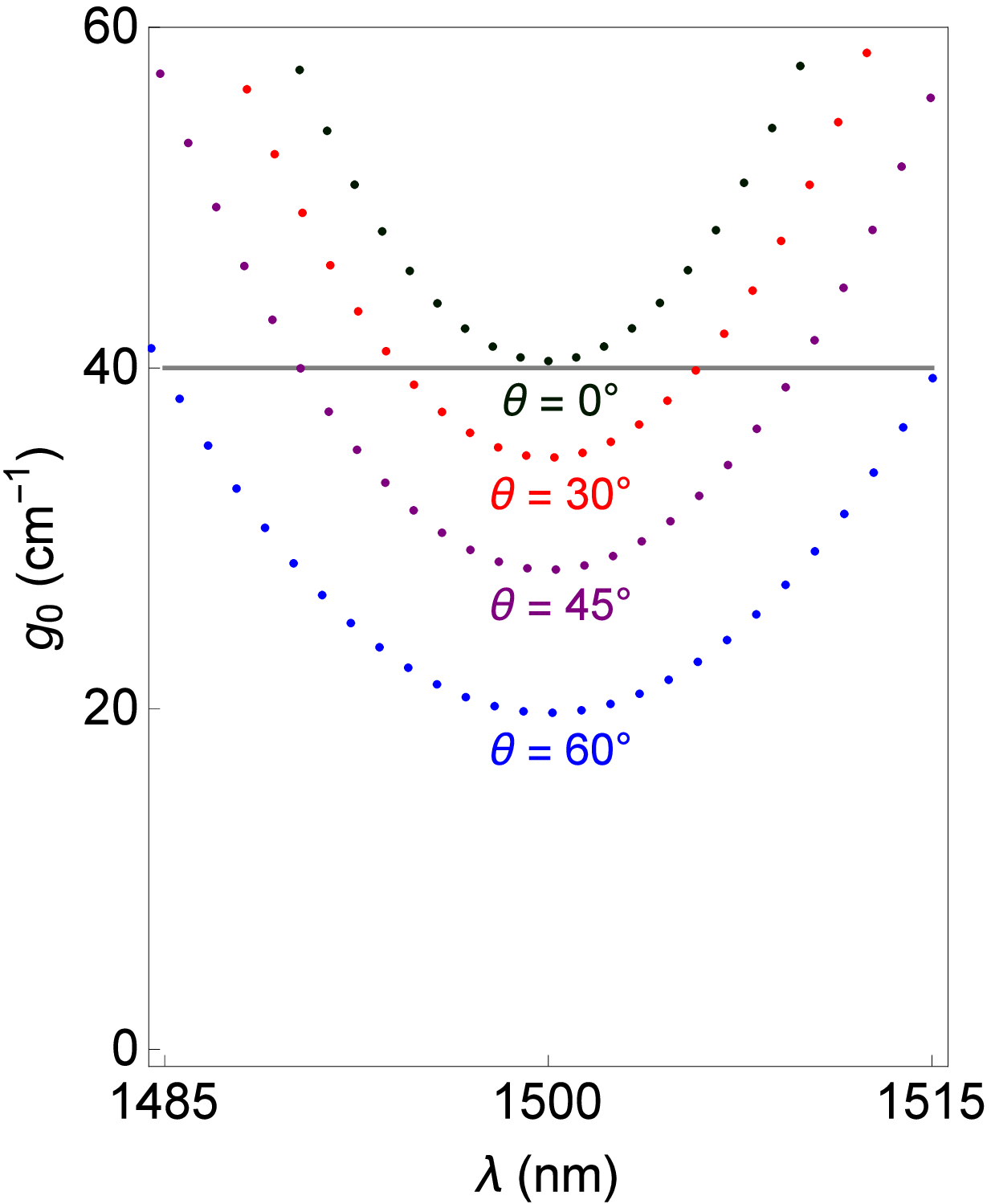}~~~~~~
    \includegraphics[scale=.50]{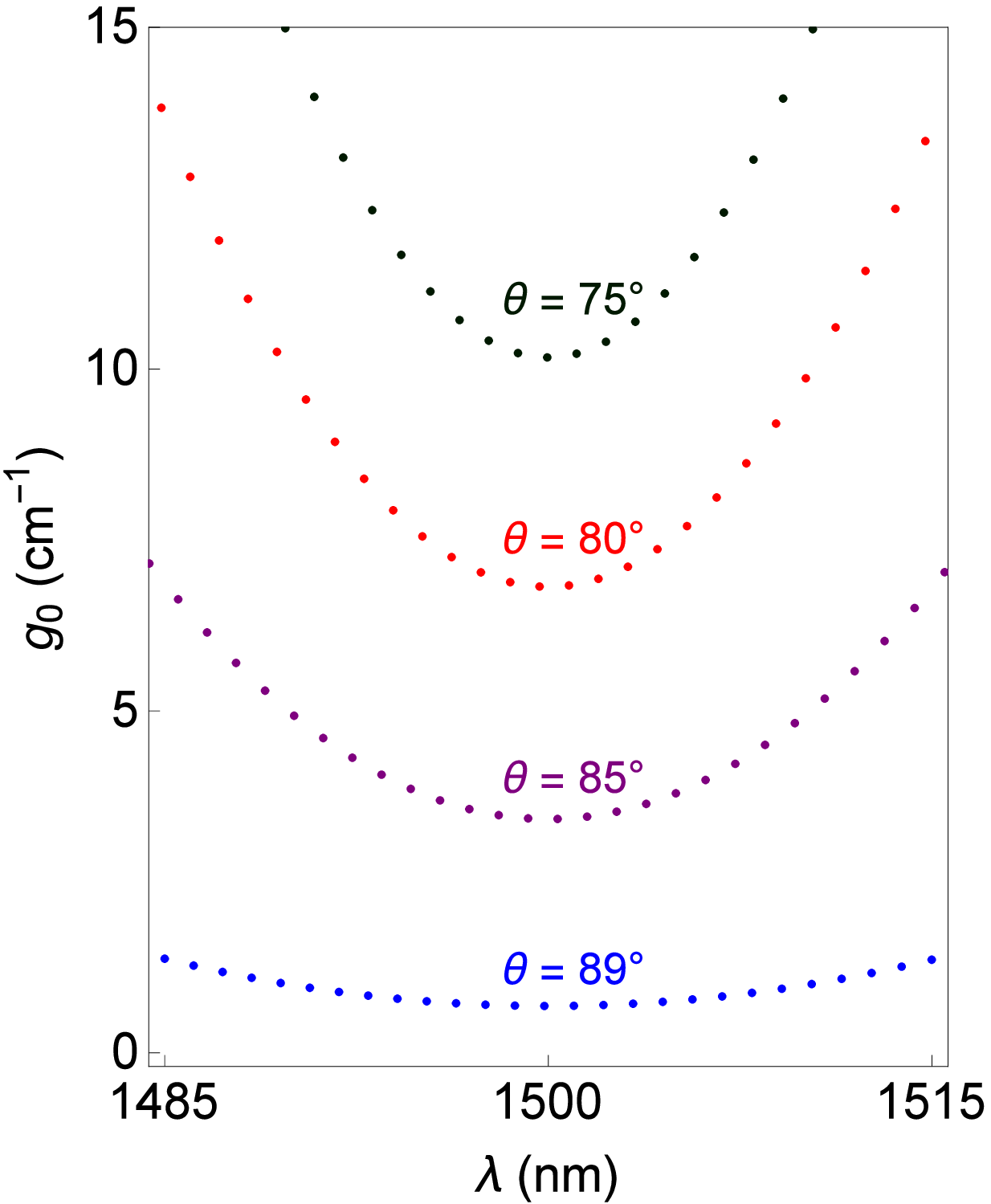}
    \caption{(Color online) Spectral singularities in the TE Modes of a planar slab of thickness $L=300~\mu m$ made out of the gain medium (\ref{specific}) for different incidence angles $\theta$. The horizontal line, $g_0=40\,{\rm cm}^{-1}$, signifies the experimental upper bound on the attainable gain.}
    \label{fig4}
    \end{center}
    \end{figure}

\begin{figure}
    \begin{center}
    \includegraphics[scale=.50]{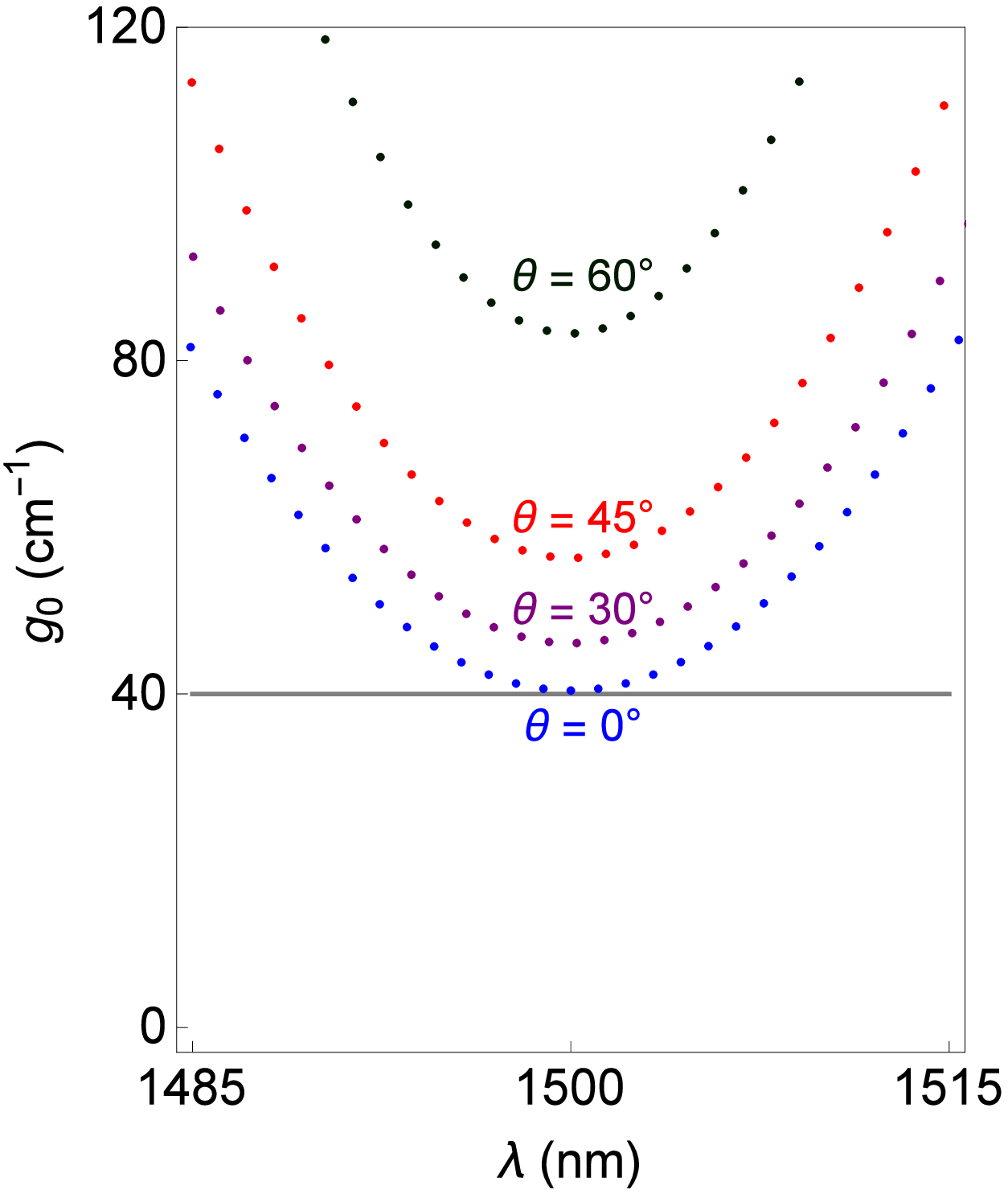}~~~~~~
    \includegraphics[scale=.50]{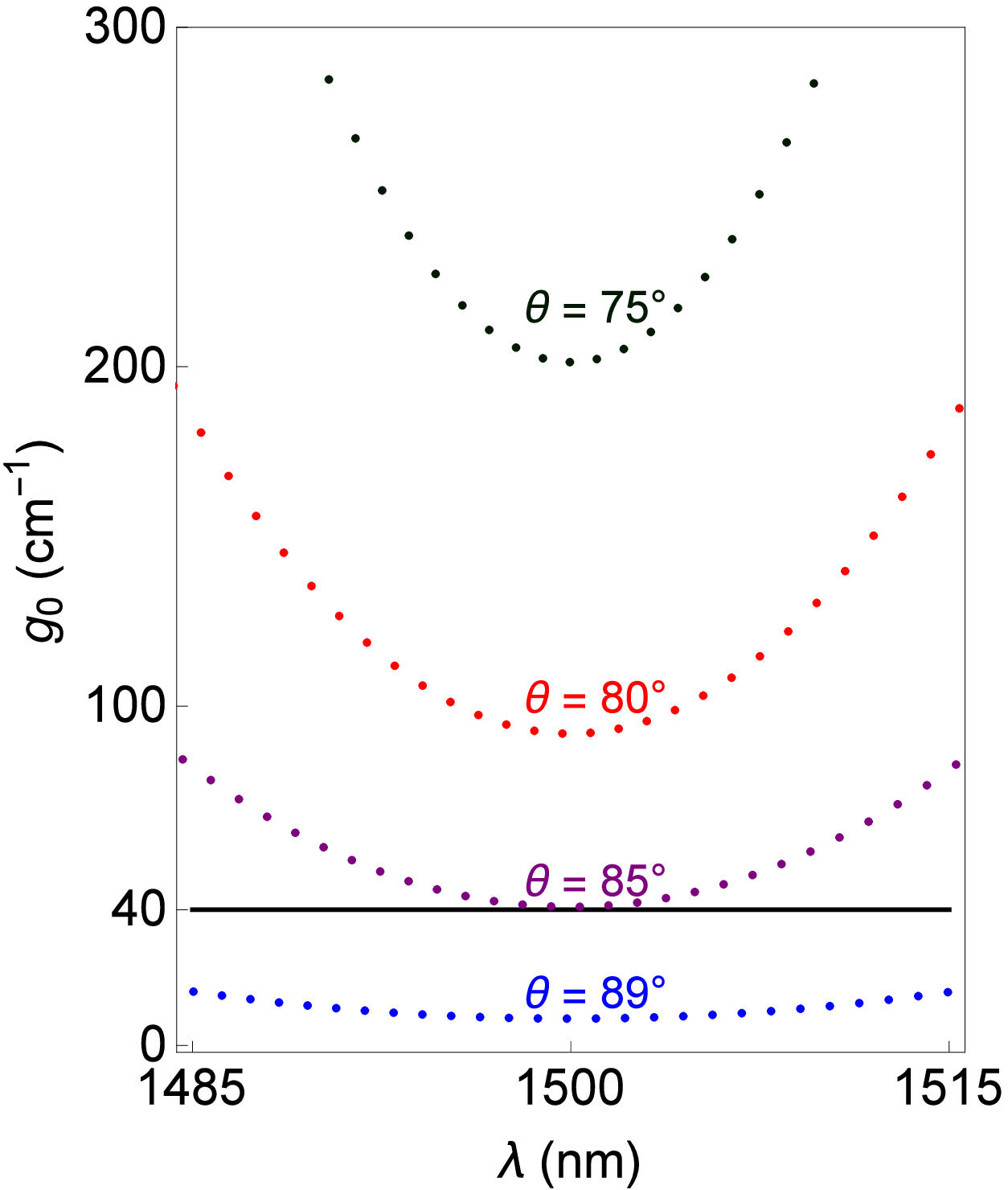}
    \caption{(Color online) Spectral singularities in the TM Modes of a planar slab of thickness $L=300~\mu m$ made out of the gain medium (\ref{specific}) for different incidence angles $\theta$. The horizontal line, $g_0=40\,{\rm cm}^{-1}$, signifies the experimental upper bound on the attainable gain.}
    \label{fig5}
    \end{center}
    \end{figure}

\begin{figure}
	\begin{center}
    \includegraphics[scale=.45]{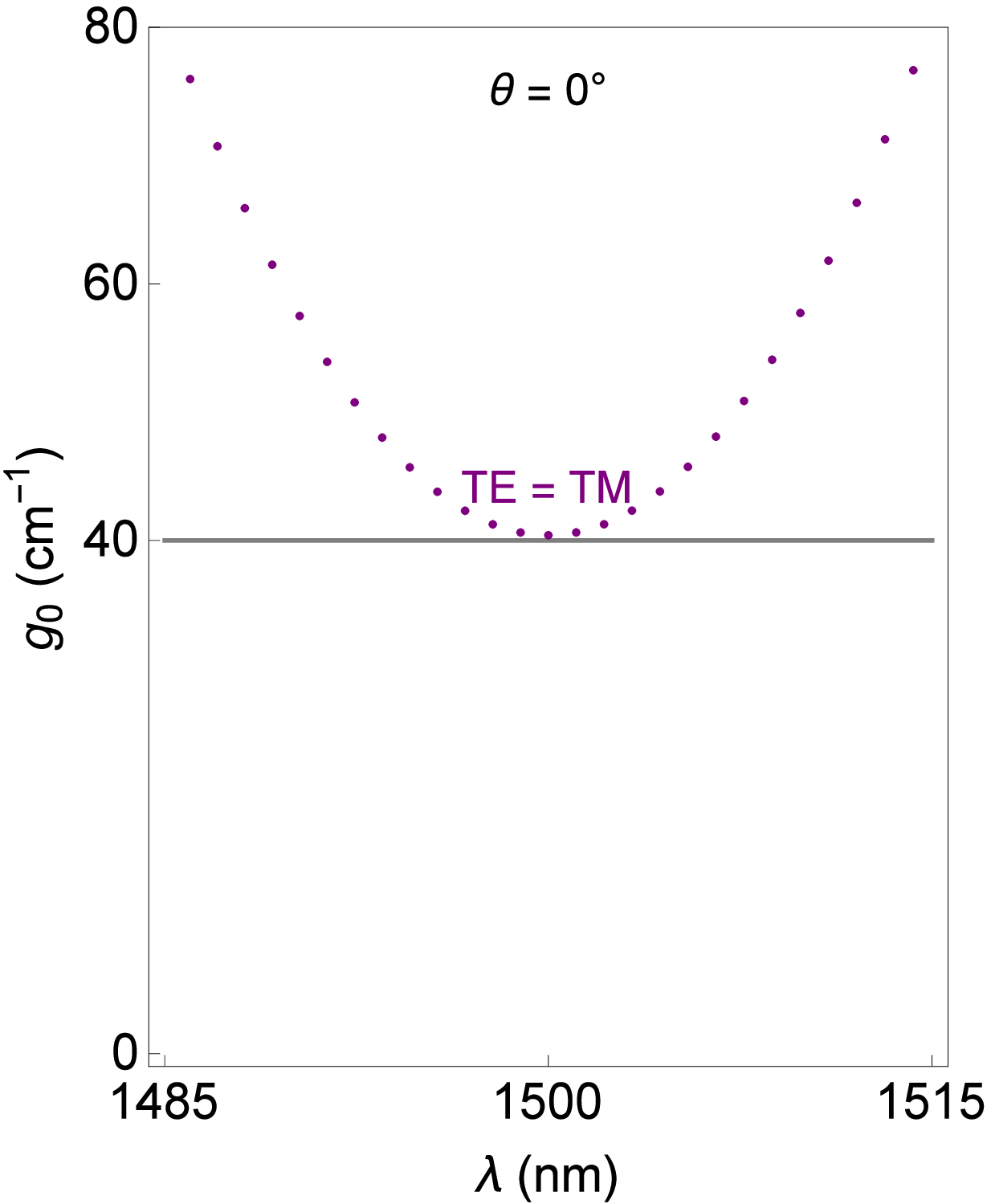}~~~
    \includegraphics[scale=.45]{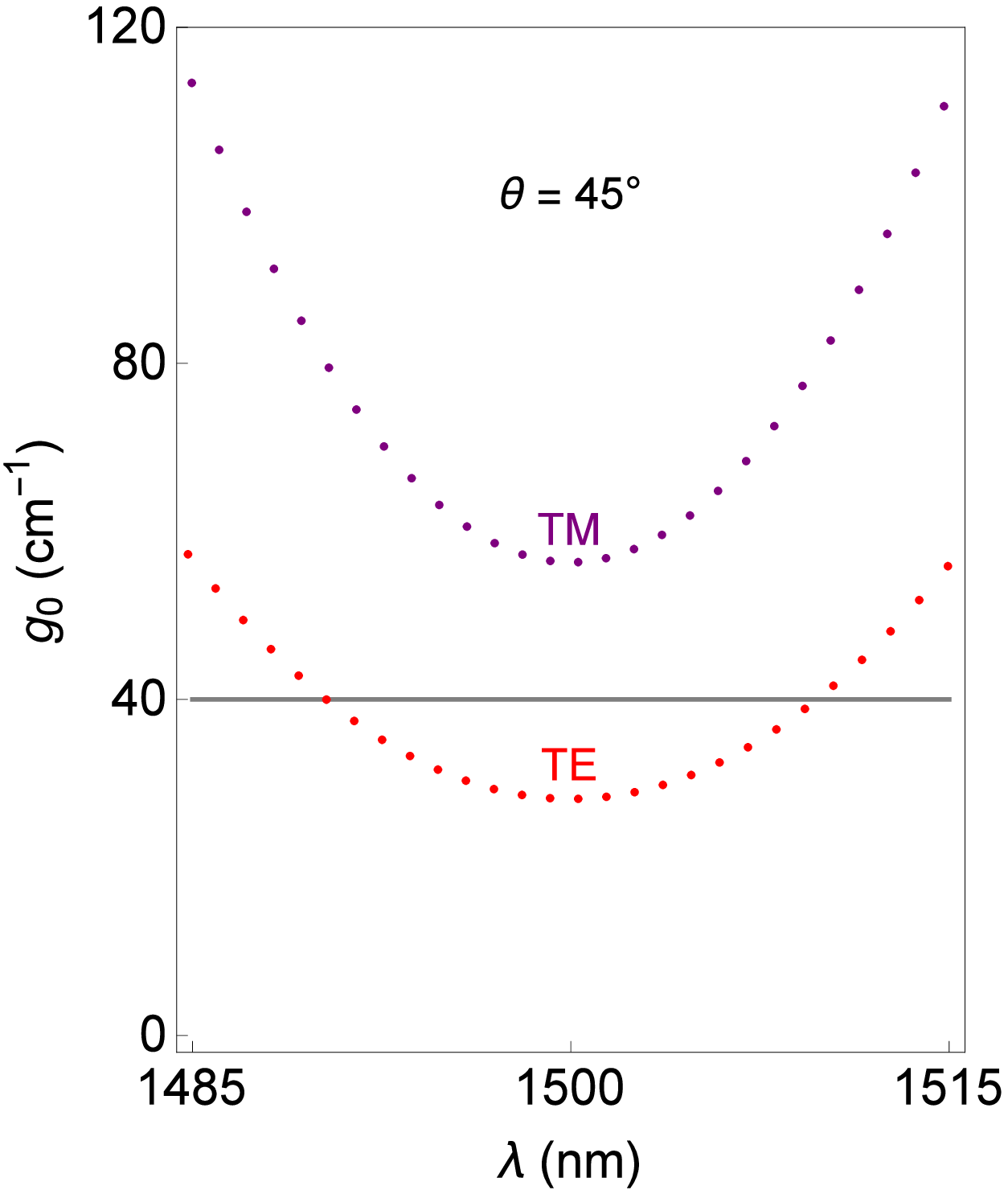}~~~
    \includegraphics[scale=.45]{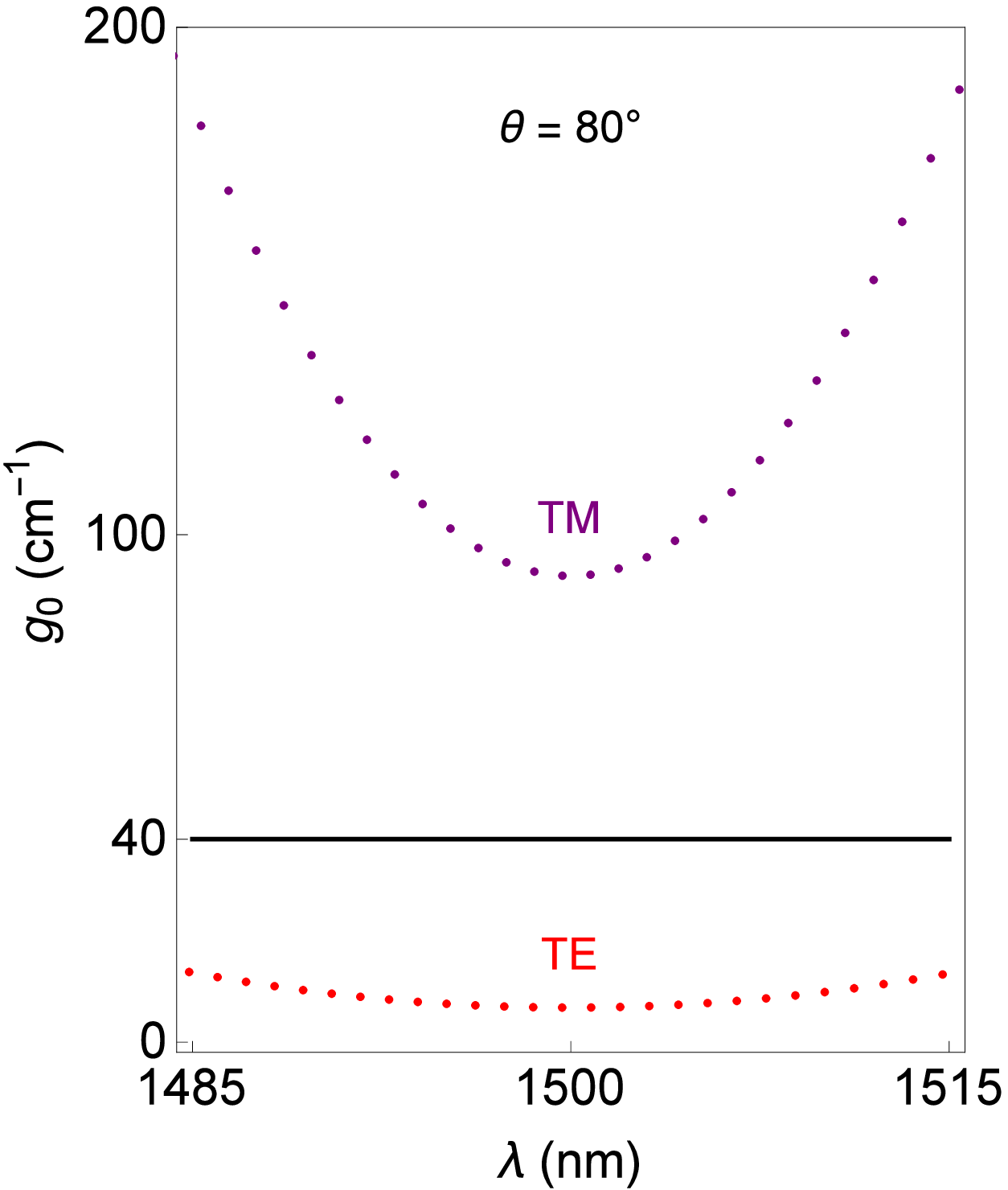}
	\caption{(Color online)  Comparison of spectral singularities in the TE and TM modes for the gain medium (\ref{specific}) and $\theta = 0^\circ, 45^\circ, 80^\circ$. The horizontal line, $g_0=40\,{\rm cm}^{-1}$, signifies the experimental upper bound on the attainable gain.}
    \label{fig6}
    \end{center}
    \end{figure}

Figures~\ref{fig4}, \ref{fig5}, and \ref{fig6} show the location of the spectral singularities in the $\lambda$-$g_0$ plane for a semiconducting slab of thickness $L=300\,\mu{\rm m}$ with the following
specifications \cite{silfvast}.
    \begin{align}
    &n_0=3.4, &&\lambda_0=1500\,{\rm nm}, &&
    \hat\gamma=0.02, &&g_0\leq 40\,{\rm cm}^{-1}.
    \label{specific}
    \end{align}
Here the last inequality gives the upper bound on the attainable gain for this material.

\section{Spectrally Singular TE and TM Waves}

If we adjust the parameters of our system so that Eq.~(\ref{normal-ss-u}) holds for a particular wavenumber $k$, the reflection and transmission amplitude diverge. This corresponds to the emergence of purely outgoing waves,
    \be
    a_0=b_2=0.
    \label{zq1}
    \ee
Equations~(\ref{normal-ss-u}) and (\ref{zq1}) together with the boundary conditions listed in Table~\ref{table02} imply
    \be
    a_2=e^{-ik_zL}b_0.
    \label{zq2}
    \ee
Furthermore, we can use (\ref{normal-ss-u}) to establish
    \be
    e^{i\tilde k z} = \left(\frac{\fu+1}{\fu-1}\right)^{\!z/L}.
    \label{ss-z}
    \ee

Substituting (\ref{zq1}) -- (\ref{ss-z}) in the formulas given in Table~\ref{table01}, we can obtain the explicit form of the spectrally singular TE and TM waves. Table~\ref{table03} gives the result of this calculation in terms of the functions,
    \begin{align}
    &F_+(\fu,z):=\left\{\begin{array}{ccc}
    e^{-ik_z z} &{\rm for} & z<0,\\
    U_+(\fu,z)/\fu &{\rm for} & 0\leq z\leq L,\\
    e^{ik_z (z-L)} &{\rm for} & z>L,\end{array}\right.
    \label{FFG-1}\\
    &F_-(\fu,z):=\left\{\begin{array}{ccc}
    e^{-ik_z z} &{\rm for} & z<0,\\
    U_-(\fu,z)&{\rm for} & 0\leq z\leq L,\\
    e^{ik_z (z-L)} &{\rm for} & z>L,\end{array}\right.
    \label{FFG-2}\\
    &G_+(\fu,z):=\left\{\begin{array}{ccc}
    e^{-ik_z z} &{\rm for} & z<0,\\
    U_+(\fu,z)/\tilde\fn&{\rm for} & 0\leq  z\leq  L,\\
    e^{ik_z (z-L)} &{\rm for} & z>L,\end{array}\right.
    \label{FFG-3}\\
    &U_\pm(\fu,z):=\frac{1}{2}\left[(\fu-1)^{1-z/L}(\fu+1)^{z/L}\pm
    (\fu-1)^{z/L}(\fu+1)^{1-z/L}\right],
    \label{Upm=}
    \end{align}
which determine the $z$-dependence of the fields. Note that the $k_x$ and $k_z$ that appear in Table~\ref{table03} and  (\ref{FFG-1}) -- (\ref{FFG-3}) are bound to take values that are consistent with Eq.~(\ref{normal-ss-u}). In particular, they satisfy
	\begin{align}
	& e^{ik_x x}=\left(\frac{\fu+1}{\fu-1}\right)^{\tan\theta\,x/\tilde n L},
	&& e^{ik_z z}=\left(\frac{\fu+1}{\fu-1}\right)^{z/\tilde n L}.
	\label{exp-x-z}
	\end{align}
    \begin{table}
    \begin{center}
	{
    \begin{tabular}{|c|c|}
    \hline
    Spectrally Singular TE-Fields & Spectrally Singular TM-Fields \\
    \hline & \\[-6pt]
    $\begin{aligned}
    & E_{x}=E_{z}=H_{y}=0\\[3pt]
    & E_{y}=b_0 e^{ik_x x}F_+( \tilde\fn,z)\\[0pt]
    &H_{x} =-\frac{b_0\cos\theta}{Z_0}\, e^{ik_x x} F_-( \tilde\fn,z)\\[0pt]
    &H_{z} =\frac{b_0\sin\theta}{Z_0}\, e^{ik_x x} F_+( \tilde\fn,z)\\[3pt]
    \end{aligned}$ &
    $\begin{aligned}
    & E_{y}=H_{x}=H_{z}=0\\[0pt]
    & E_{x} =b_0 Z_0\cos\theta\,  e^{ik_x x} F_-(\frac{\tilde\fn}{\fn^2},z)\\[0pt]
    & E_{z} =- b_0 Z_0\sin\theta \, e^{ik_x x}G_+(\frac{\tilde\fn}{\fn^2},z)\\[0pt]
    & H_{y} =b_0 e^{ik_x x} F_+( \frac{\tilde\fn}{\fn^2},z)\\[-8pt]
    &
    \end{aligned}$\\
    \hline
    \end{tabular}}
    \vspace{6pt}
    \caption{Components of the spectrally singular TE and TM fields in cartesian coordinates. Here $F_\pm(\fu,z)$ and $G_+(\fu,z)$ are defined by (\ref{FFG-1}) -- (\ref{FFG-3}), and (\ref{exp-x-z}) holds.}
    \label{table03}
    \end{center}
    \end{table}%
The transformation properties of $F_\pm, G_+$, and $U_\pm$ under the parity transformation,
	\be
	z\stackrel{\cP}{\longrightarrow} L-z,
	\label{parity}
	\ee
provide a direct verification of the $\cP$-invariance of the phenomenon of spectral singularities. This is important, because it is the basic symmetry responsible for the coherent emission of waves from the left and right boundaries of the slab at a spectral singularity \cite{pra-2013c}. We provide a more explicit demonstration of this behavior in the following section where we examine the Poynting vector for the spectrally singular waves.

\section{Poynting Vector and Energy Density at a Spectral Singularity}
\label{section-5}

If the slab we consider has the gain coefficient $g$ required for generating a spectral singularity in an oblique TE or TM mode, it begins lasing in this mode. In this section we examine the behavior of the time-averaged Poynting vector and the energy density inside and outside the slab for the values of the physical parameters of the slab that realize such a spectral singularity. The time-averaged Poynting vector and the energy density are respectively given by    		
\cite{jackson}
    \begin{align}
    &\langle\vec{S}\rangle = \frac{1}{2}\,\RE \left(\vec{E} \times \vec{H}^*\right),
    \label{S-def}\\
    &\br u\kt := \frac{1}{4}\RE\left(\vec{E}\cdot \vec{D}^* +\vec{B}\cdot \vec{H}^*\right)=
    \frac{1}{4}\left(\epsilon_0\RE[\fz(z)]|\vec{E}|^2 +\mu_0|\vec{H}|^2\right).
    \label{u-def}
    \end{align}
We calculate these quantities for the spectrally singular modes of our system by substituting the formulas given in Table~\ref{table03} in Eqs.~(\ref{S-def}) and (\ref{u-def}). This yields
    \bea
    \langle\vec{S}^{(E/M)}\rangle&=&
    |\br\vec S_{\rm out}^{(E/M)}\kt|\times \left\{\begin{array}{ccc}
    \sin\theta\, \hat e_x-\cos\theta\,\hat e_z & {\rm for} & z<0,\\[3pt]
    \cX^{(E/M)}(z) \sin\theta\,\hat e_x+\cZ^{(E/M)}(z)\cos\theta\,\hat e_z
     & {\rm for} & 0\leq z\leq L,\\[3pt]
    \sin\theta\,\hat e_x+\cos\theta\, \hat e_z & {\rm for} & z> L,
    \end{array}\right.
    \label{EM-poynting}\\[6pt]
    \br u^{(E/M)}\kt &=& \br u^{(E/M)}_{\rm out}\kt
    \times\left\{
    \begin{array}{ccc}
    \cU^{(E/M)}(z) & {\rm for} & 0\leq z\leq L,\\[3pt]
    1 & {\rm for} & z<0~{\rm and}~z>L,\end{array}\right.
    \label{EM-u}
    \eea
where the superscripts $(E)$ and $(M)$ refer to the TE and TM waves, and
	\begin{align}
	&\begin{aligned}
	&|\br\vec S_{\rm out}^{(E)}\kt|:=\frac{|b_0|^2}{2Z_0},
	&&|\br\vec S_{\rm out}^{(M)}\kt|:=\frac{Z_2|b_0|^2}{2},\nn\\
	&\cX^{(E)}(z):=\left|\frac{U_+(\tilde\fn,z)}{\tilde\fn}\right|^2,
	&& \cX^{(M)}(z):=\left|\frac{U_+(\tilde\fn/\fn^2,z)}{\tilde\fn}\right|^2\RE(\fn^2),\nn\\
	&\cZ^{(E)}(z):=\RE\left[\frac{U_+(\tilde\fn,z)U_-(\tilde\fn,z)^*}{\tilde\fn}\right],
	&& \cZ^{(M)}(z):=\RE\left[\frac{\fn^2U_+
	(\tilde\fn/\fn^2,z)U_-(\tilde\fn/\fn^2,z)^*}{\tilde\fn}\right],\nn\\
	&\br u^{(E)}_{\rm out}\kt:=\frac{\epsilon_0|b_0|^2}{2},
	&&\br u^{(M)}_{\rm out}\kt:=\frac{\mu_0|b_0|^2}{2},\nn
	\end{aligned}\nn\\
	&\begin{aligned}
	&\cU^{(E)}(z):=\frac{1}{2|\tilde\fn|^2}\left\{
	\left[\RE(\fn^2)+\sin^2\theta\right]
	\left|U_+(\tilde\fn,z)\right|^2+
	\cos^2\theta\,\left|\tilde\fn\,U_-(\tilde\fn,z)\right|^2\right\},\nn\\
	&\cU^{(M)}(z):=\frac{1}{2|\tilde\fn|^2}\left\{
	\left[|\fn|^4+\sin^2\theta\,\RE(\fn^2)\right]
	\left|U_+(\tilde\fn/\fn^2,z)\right|^2+
	\cos^2\theta\,\RE(\fn^2)\left|\tilde\fn\,U_-(\tilde\fn/\fn^2,z)\right|^2\right\}.\nn
	\end{aligned}\nn
	\end{align}
	
\begin{figure}
    \begin{center}
    \includegraphics[scale=.60]{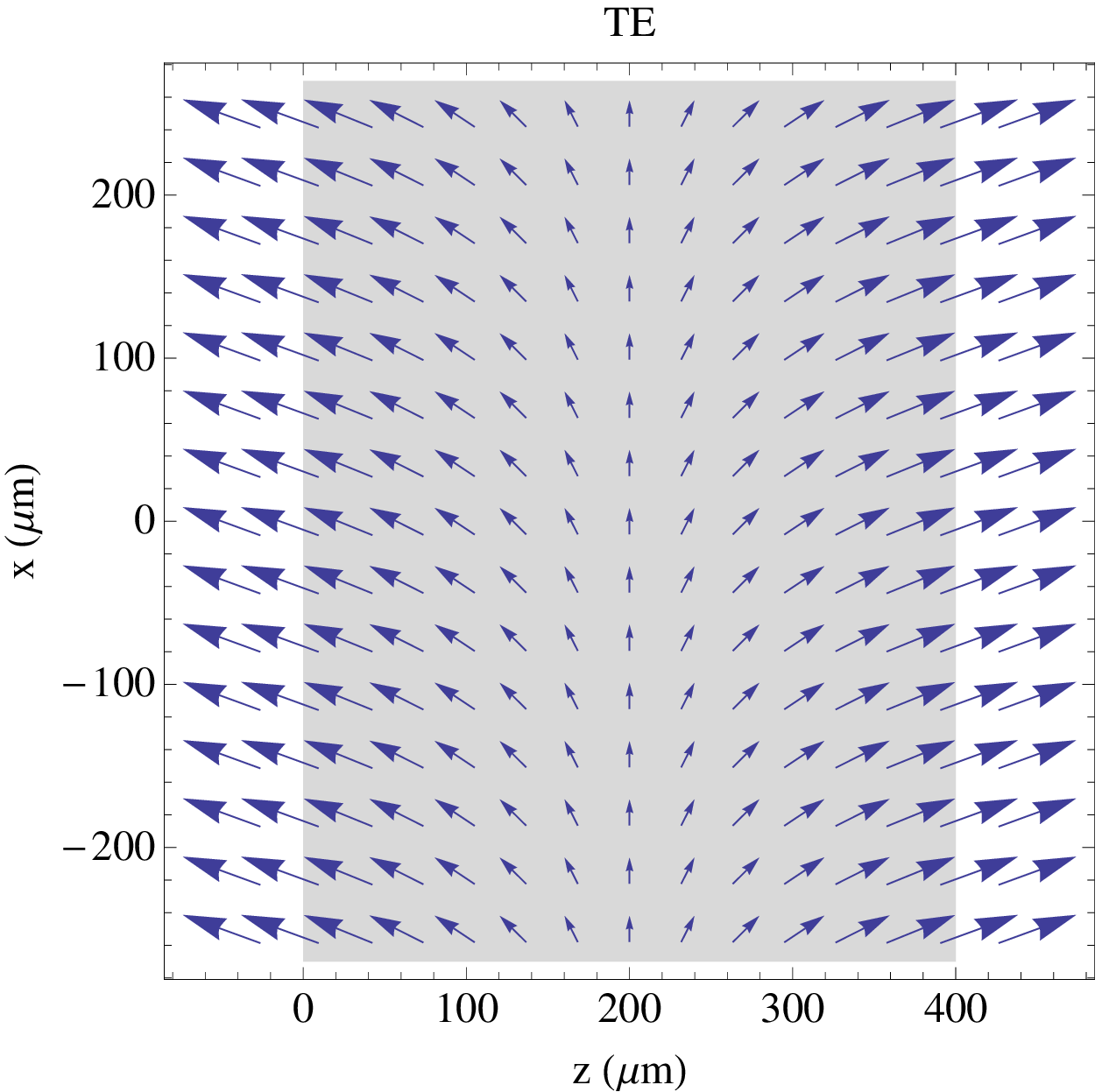}~~~~~~
    \includegraphics[scale=.60]{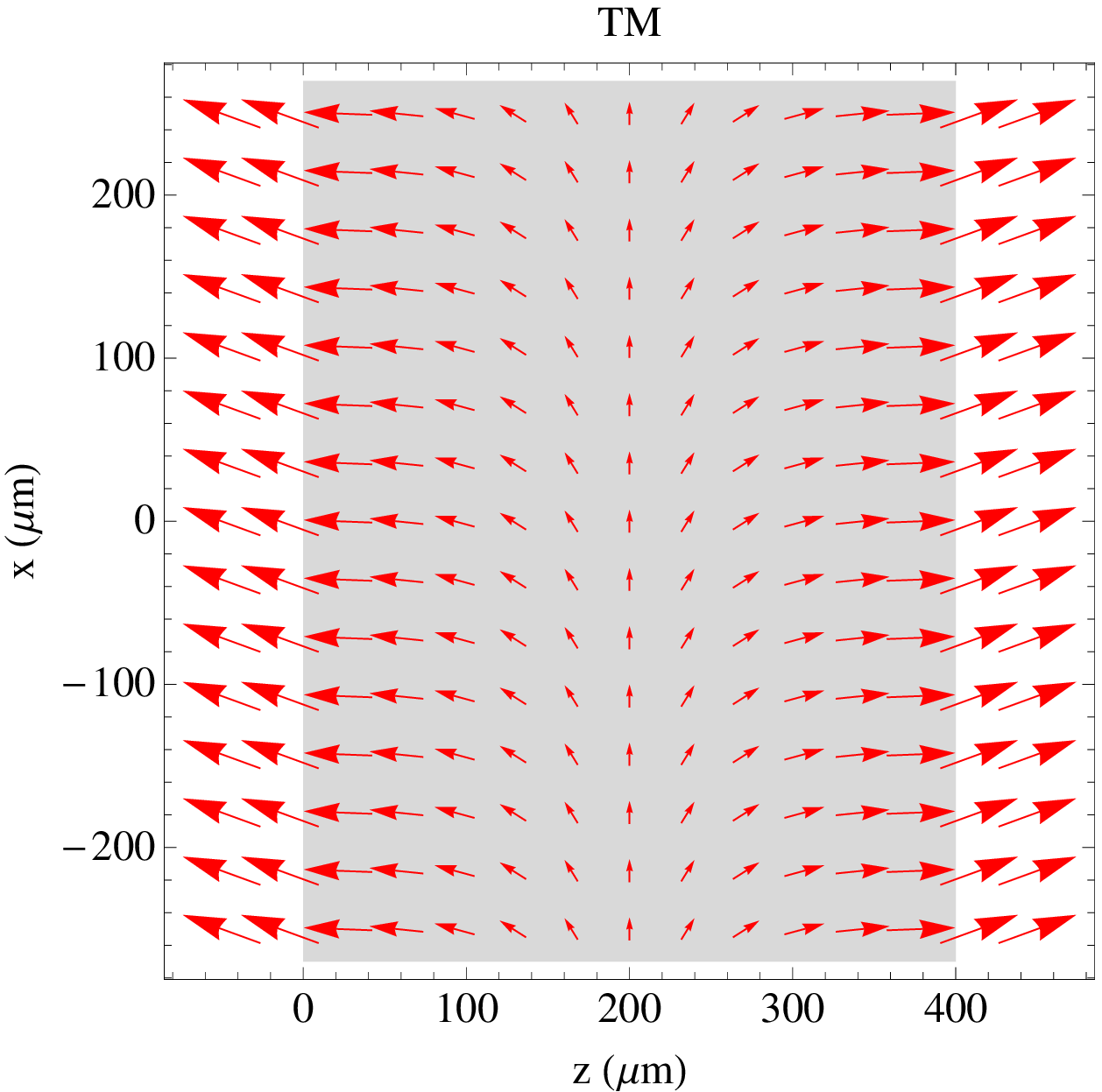}
    \caption{(Color online) Graphs of the time-averaged Poynting vectors $\langle\vec{S}^{(E)}\rangle$ (on the left in blue) and $\langle\vec{S}^{(M)}\rangle$ (on the right in red) for spectral singularities in oblique TE and TM modes with $\theta=20^\circ$ and
    specifications~(\ref{SS-20=}). Here $L=400\,\mu{\rm m}$ and $\eta=3.4$. The gray region represents the slab.}
    \label{fig7}
    \end{center}
    \end{figure}

\begin{figure}
    \begin{center}
    \includegraphics[scale=.5]{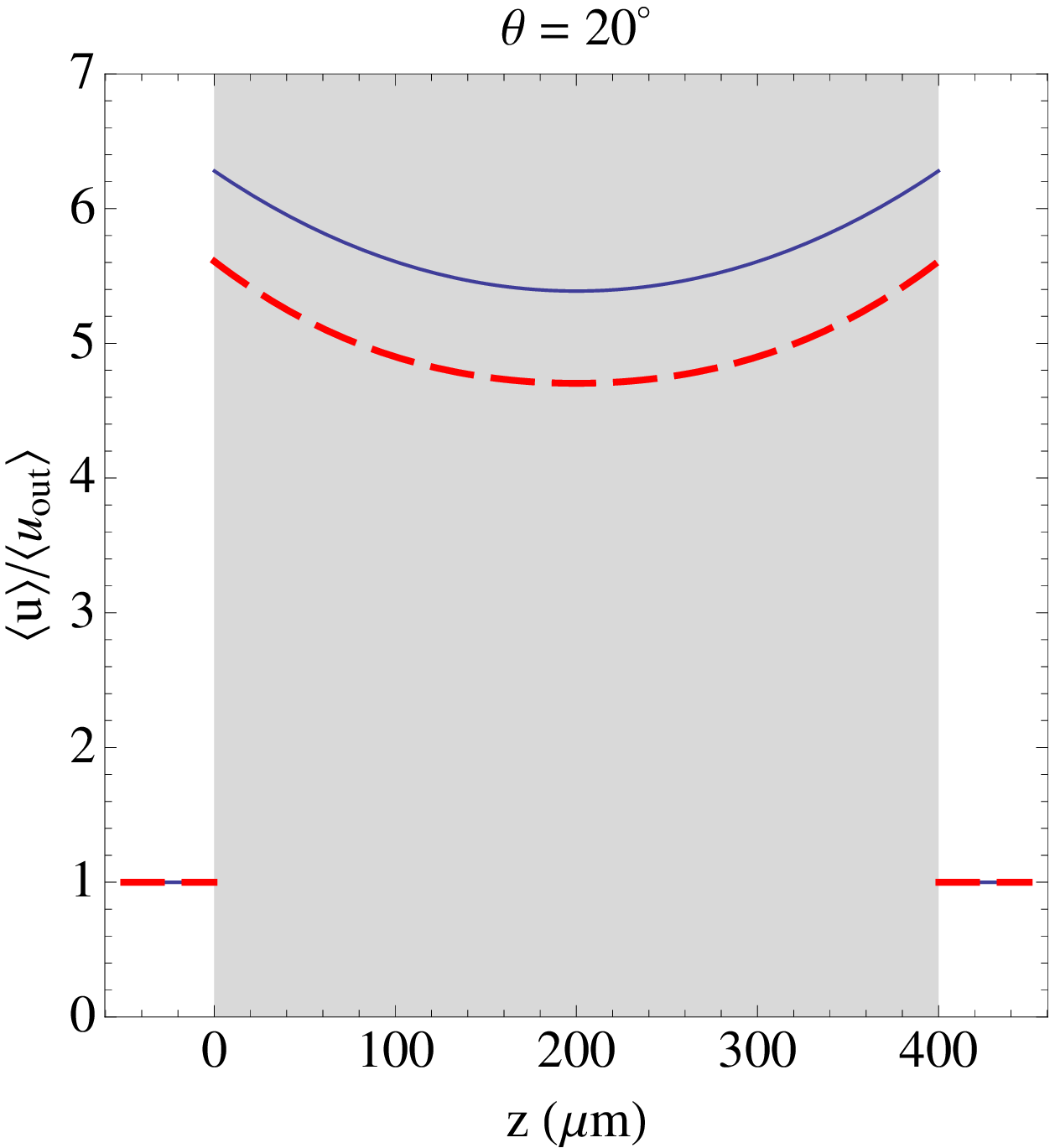}~~~~~~~~~
    \includegraphics[scale=.5]{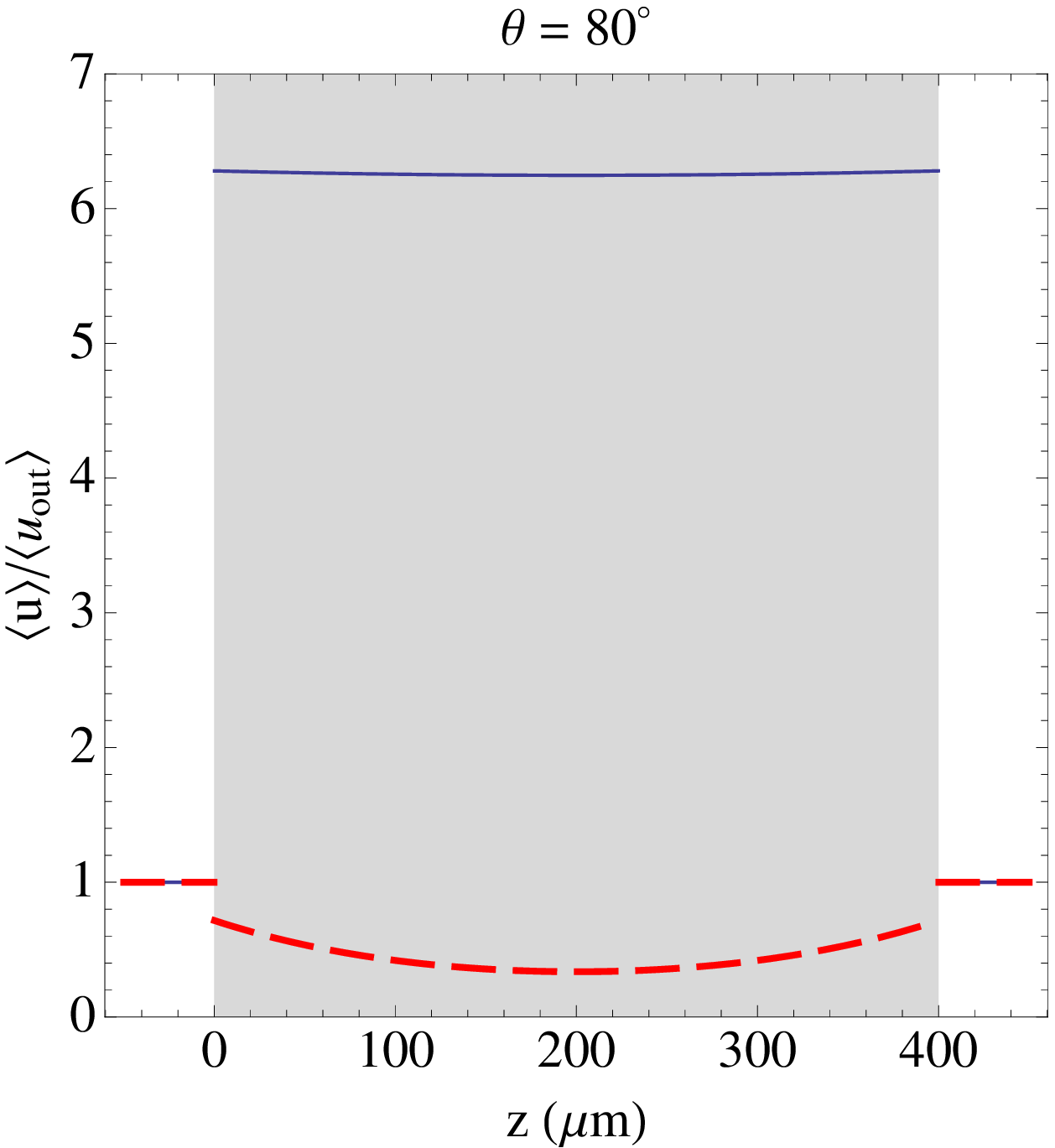}
    \caption{(Color online) Graphs of the normalized energy density $\langle u \rangle/\langle u_{\rm out}\rangle$ for the spectrally singular TE (solid blue curves) and TM (dashed red curves) modes. The left-hand figure corresponds to the configuration depicted in Fig.~\ref{fig7} that has $\theta=20^\circ$. The right-hand figure corresponds to spectral singularities (\ref{SS-80=}) with the same values of $L$ and $\eta$, but for $\theta=80^\circ$. Again the gray area corresponds to the gain region.}
    \label{fig8}
    \end{center}
    \end{figure}

Figure~\ref{fig7} shows plots of $\langle\vec{S}\rangle$ for a slab of thickness $L=400\mu{\rm m}$ and real refractive index $\eta=3.4$ that admits spectrally singular TE and TM modes with $\theta=20^\circ$ and
    \begin{align}
    &\lambda^{(T/E)}=1500.111~{\rm nm}, && g^{(E)}=28.385\,{\rm cm}^{-1},
    && g^{(M)}=32.048\,{\rm cm}^{-1}.
    \label{SS-20=}
    \end{align}
Figure~\ref{fig8} shows the plots of $\langle u \rangle/\langle u_{\rm out}\rangle$ for the spectrally singular modes of Fig.~\ref{fig7} and spectrally singular TE and TM modes with $\theta=80^\circ$ and the following specifications
    \be
    \begin{aligned}
    &\lambda^{(E)}=1500.519~{\rm nm}, && g^{(E)}=5.118\,{\rm cm}^{-1},\\
    &\lambda^{(M)}=1500.506~{\rm nm}, && g^{(M)}=68.982\,{\rm cm}^{-1}.
    \label{SS-80=}
    \end{aligned}
    \ee
These respectively provide the general behavior of $\langle u \rangle$ for $\theta<\theta_c$ and $\theta>\theta_c$. In particular, for the TM modes with $\theta>\theta_c$ the energy density inside the slab is smaller than that of outside the slab. This is in sharp contrast to TE modes and TM modes with $\theta<\theta_c$. The same applies to the magnitude of time-averaged Poynting vector; for TM modes with $\theta>\theta_c$, $|\br \vec S\kt|$ takes much smaller values inside the slab and increases sharply at the boundaries.

Figures~\ref{fig7} and \ref{fig8} provide a clear demonstration  of the $\cP$-symmetry of the phenomenon of spectral singularities. It is indeed easy to establish the existence of this symmetry using Eqs.~(\ref{EM-poynting}) and (\ref{EM-u}). Under the parity transformation (\ref{parity}), $U_\pm(\fu,z)$, $\cX^{(E/M)}$, $\cZ^{(E/M)}$, and $\cU^{(E/M)}$ transform according to
	\begin{align}
	&U_\pm(\fu,z)\to \pm U_\pm(\fu,z), && \cX^{(E/M)}\to\cX^{(E/M)}, \nn\\
	&\cZ^{(E/M)}\to -\cZ^{(E/M)}, && \cU^{(E/M)}\to \cU^{(E/M)}.\nn
	\end{align}
These in turn imply that $\langle\vec{S}^{(E/M)}\rangle$ and $\br u^{(E/M)}\kt$ are $\cP$-invariant quantities.

The following are some of the other notable features of $\langle\vec{S}^{(E/M)}\rangle$ and $\br u^{(E/M)}\kt$ that can be established using the analytic expressions given in (\ref{EM-poynting}) and (\ref{EM-u}) or seen from Figs.~\ref{fig7} and \ref{fig8}.
	\begin{enumerate}
	\item $\langle\vec{S}^{(E)}\rangle$ is a continuous function of $z$. In particular, if we denote the angle between $\langle\vec{S}^{(E)}\rangle$ and the positive $z$-axis by $\Theta^{(E)}(z)$, we find that $\Theta^{(E)}(0)=\pi-\theta$ and  $\Theta^{(E)}(L)=\theta$.
	\item $\langle\vec{S}^{(M)}\rangle$ is a continuous function of $z$ except at the boundaries of the slab where its $x$-component undergoes jumps. This follows from (\ref{EM-poynting}) and the fact that
		\bea
		\cX^{(M)}(0)&=&\cX^{(M)}(L)=\frac{\RE(\fn^2)}{|\fn|^4}\neq 1,\nn\\
		\cZ^{(M)}(0)&=&-\cZ^{(M)}(L)=-1.\nn
		\eea
In particular, the angle $\Theta^{(M)}(z)$ between $\langle\vec{S}^{(M)}\rangle$ and the positive $z$-axis satisfy
	\begin{align*}
    &\Theta^{(M)}(z)=\left\{\begin{array}{ccc}
	\pi-\theta &{\rm for} & z<0,\\
	\pi-\tilde\theta&{\rm for} & z=0,\\
	\tilde\theta&{\rm for} & z=L,\\
	\theta &{\rm for} & z>L,
	\end{array}\right. && \tilde\theta:=\tan^{-1}\left[\frac{\tan\theta\,\RE(\fn^2)}{|\fn|^4}\right].
    \end{align*}
For the spectrally singular TM mode depicted in Fig.~\ref{fig7} where $\theta=20^\circ$, we have $\tilde\theta=1.8^\circ$.

\item At $z=L/2$ both $\langle\vec{S}^{(E)}\rangle$ and $\langle\vec{S}^{(M)}\rangle$ point along the positive $x$-axis, except for $\theta=0$ where they vanish identically;
	\bea
	\left.\langle\vec{S}^{(E)}\rangle\right|_{z=L/2}&=&|\langle\vec{S}_{\rm out}^{(E)}\rangle|
	\left|\tilde\fn-\tilde\fn^{-1}\right|^2\sin\theta\,\hat e_x,\nn\\
	\left.\langle\vec{S}^{(M)}\rangle\right|_{z=L/2} &=&|\langle\vec{S}_{\rm out}^{(M)}\rangle|
	\left(\frac{\RE(\fn^2)}{|\fn|^4}\right)\left|\frac{\tilde\fn}{\fn^2}-
	\frac{\fn^2}{\tilde\fn}\right|^2\sin\theta\,\hat e_x.\nn
	\eea

\item Inside the slab $|\langle\vec{S}^{(E/M)}\rangle|$ and $\br u^{(E/M)}\kt$ attain their minimum value at $z=L/2$. The latter grows monotonically as one approaches the boundary.

	\end{enumerate}

The graphical description of  $\langle\vec{S}^{(E/M)}\rangle$ and $\br u^{(E/M)}\kt$ for spectral singularities with different values of physical parameters turns out to have general validity. In particular, there is no qualitative difference between oblique spectrally singular TE modes with different $\theta$. For the oblique spectrally singular TM modes there are basic differences of behavior between the modes with $\theta<\theta_b$ and those with $\theta>\theta_b$.

\section{Concluding Remarks}

In this article we have used the recent developments regarding the optical realizations of spectral singularities to provide a mathematically precise derivation of the laser threshold condition for the oblique TE and TM modes of an infinite planar slab. The standard textbook treatments of this problem rely on the principle of conservation of energy \cite{silfvast}. Our approach has the advantage of not relying on any physical principles but only on the Maxwell's equations and the definition of a spectral singularity.

For the TE waves, the essential features of spectral singularities, which determine the lasing threshold and the CPA conditions, are not sensitive to the direction of the propagation of the wave. This turns out not to be the case for the TM waves. If the incidence (or rather emission) angle $\theta$ is smaller than the Brewster's angle $\theta_b$, the spectrally singular TM waves behave exactly like spectrally singular TE waves, except that their threshold gain is an increasing function of $\theta$. For the TM waves with $\theta>\theta_b$ the situation is completely different.

A basic feature of spectral singularities is that they are $\cP$-invariant. This symmetry is responsible for the fact that at a spectral singularity the emitted waves from the left  and right  boundaries of the slab have identical amplitude and phase, i.e., the slab emits coherent waves. We have provided a detailed examination of this phenomenon by offering explicit formulas for the energy density and Poynting vector of the oblique spectrally singular TE and TM waves. This reveals the curious fact that the energy density of a spectrally singular TM wave with $\theta>\theta_b$ is smaller inside the gain region than the surrounding vacuum. This is the opposite of what happens for spectrally singular TM waves with $\theta<\theta_b$ and the spectrally singular TE waves with arbitrary $\theta$. This observation has counterintuitive implications. For example it means that if we use a lossy material and arrange that its attenuation coefficient has the same value as the threshold gain coefficient supporting an oblique spectrally singular TM mode with $\theta>\theta_b$, the slab will function as a CPA having a lower energy density inside, i.e., the waves are absorbed more predominantly by its boundary.\\[6pt]

\noindent{\em Acknowledgments:}  We are grateful to Ali Serpeng\"{u}zel for fruitful discussions. This work has been supported by  the Scientific and Technological Research Council of Turkey (T\"UB\.{I}TAK) in the framework of the project no: 112T951, and by the Turkish Academy of Sciences (T\"UBA).

\end{document}